\documentclass[conference]{IEEEtran}
\IEEEoverridecommandlockouts
\usepackage{caption} % Add this at the top of your LaTeX document
\usepackage{cite}
\usepackage{amsmath,amssymb,amsfonts}
\usepackage{algorithmic}
\usepackage{graphicx}
\usepackage{textcomp}
\usepackage{xcolor}
\usepackage{hyperref} % Required for clickable URLs
\usepackage{graphicx} % Required for image inclusion
\usepackage{epsfig} % Alternative package for handling .ps and .ep
\def\BibTeX{{\rm B\kern-.05em{\sc i\kern-.025em b}\kern-.08em
    T\kern-.1667em\lower.7ex\hbox{E}\kern-.125emX}}
    
\begin{document}

\title{HERMES: High-Performance RISC-V Memory Hierarchy for ML Workloads

}

\author{
    \IEEEauthorblockN{Pranav Suryadevara}
    \IEEEauthorblockA{
        Morris Plains, NJ, USA \\
        pranavsuryadevara@yahoo.com
    }
}

\maketitle

\begin{abstract}
The growth of machine learning (ML) workloads has underscored the importance of efficient memory hierarchies to address bandwidth, latency, and scalability challenges. HERMES focuses on optimizing memory subsystems for RISC-V architectures to meet the computational needs of ML models such as CNNs, RNNs, and Transformers. This project explores state-of-the-art techniques such as advanced prefetching, tensor-aware caching, and hybrid memory models. The cornerstone of HERMES is the integration of shared L3 caches with fine-grained coherence protocols equipped with specialized pathways to deep-learning accelerators such as Gemmini. Simulation tools like Gem5 and DRAMSim2 were used to evaluate baseline performance and scalability under representative ML workloads. The findings of this study highlight the design choices, and the anticipated challenges, paving the way for low-latency scalable memory operations for ML applications.
\end{abstract}

\bigskip

\begin{IEEEkeywords}
RISC-V, Memory Hierarchy, Machine Learning, Shared L3 Cache, Tensor-Aware Caching, Advanced Prefetching, Gemmini, Gem5, DRAMSim2, ML Workloads, Memory Bottlenecks.
\end{IEEEkeywords}

\section{Introduction}

Machine learning (ML) models have become integral to modern computing, powering applications across fields such as Computer Vision, Natural Language Processing, and Autonomous Systems. With these advancements, models such as Convolutional Neural Networks (CNNs), Recurrent Neural Networks (RNNs), Transformers, Large Language Models (LLMs), and Large Vision-Language Models (LVLMs) often require extensive computational resources and memory bandwidth. Hence as a result optimizing memory hierarchies for ML workloads is of significant importance for enhancing performance and efficiency \cite{jouppi2017, chen2019}.

RISC-V, an open-source Instruction Set Architecture (ISA) initially developed at the University of California, Berkeley, is gaining traction due to its flexibility and extensibility \cite{patterson2017, asanovic2014}. However, the performance of RISC-V processors for ML workloads is currently limited by memory bottlenecks, including latency, limited bandwidth, and lack of scalability \cite{manchala2024}. The HERMES project aims to design a high-performance memory subsystem that is tailored to optimize ML workloads on RISC-V architectures. Using techniques such as advanced prefetching \cite{ganfure2020}, tensor-aware caching \cite{seshadri2014}, and hybrid memory models \cite{rang2024}, HERMES seeks to overcome these challenges and enhance the efficiency of ML computations.

This report presents a foundational study of HERMES, outlining the design considerations involved, the initial simulation results obtained, and the challenges anticipated to achieve a scalable, low-latency memory hierarchy system for ML workloads. The project evaluates the performance of the system using simulation tools such as Gem5 \cite{binkert2011} and DRAWSim2\cite{drawsim}.

\section{Literature Survey}

\subsection{Memory Bottlenecks in ML Workloads}

ML workloads, especially the deep-learning models, are characterized by large datasets and frequent memory access patterns. Studies have identified that the key challenges that impact the efficiency of memory subsystems are:

\subsubsection*{High Latency}

Tensor operations, such as matrix multiplication, require frequent access to large datasets stored in memory. The latency factor involves fetching in data from DRAM which can significantly degrade performance if it isn't mitigated by efficient caching and prefetching mechanisms \cite{jouppi2017, chen2019, chishti2019}.

\subsubsection*{Bandwidth Constraints}

ML models, particularly those involving convolutions and recurrent operations, demand a high memory bandwidth. The concurrent fetching of weights, activations, and gradients often saturates the available bandwidth, leading to bottlenecks in the data transfer. This issue is exacerbated in systems with limited memory available on the chip \cite{manchala2024, rang2024, ali2022}.

\subsubsection*{Scalability}

As ML models grow in size and complexity, memory subsystems must scale to accommodate the increasing data loads. Current architectures struggle to maintain performance as the size of the dataset exceeds the cache capacity, thus requiring frequent access to the slower off-chip memory \cite{binkert2011, devarapu2021}.

Addressing these bottlenecks requires innovative memory hierarchies that can balance latency, bandwidth, and scalability for a variety of ML workloads.

\subsection{Existing Memory Solutions for ML}
Various techniques have been developed to optimize memory systems for ML workloads. These solutions address specific aspects of memory bottlenecks and have been implemented in modern hardware architectures.

\subsubsection*{Tensor-Aware Caching}

Tensor-aware caching improves data reuse by optimizing the cache layout for tensor operations. For example, Google’s Tensor Processing Unit (TPU) uses specialized caching mechanisms that align with the structure of tensor computations, reducing latency and increasing throughput \cite{jouppi2017}. This approach ensures that frequently accessed tensors remain in fast on-chip memory, minimizing DRAM accesses.

\subsubsection*{Prefetching Strategies}

Advanced prefetching techniques predict data access patterns and load data into the cache before it is needed. Techniques such as Stride Prefetching and Machine Learning-Based Prefetching have been proven effective in reducing latency for sequential and non-sequential data access patterns \cite{ganfure2020, chishti2019}. For instance, Intel’s microprocessors incorporate sophisticated hardware prefetchers that dynamically adapt to the workload behavior at hand \cite{seshadri2014, ali2022}.

\subsubsection*{Hybrid Memory Models}

Combining DRAM with High-Bandwidth Memory (HBM) offers a balance between capacity and speed. HBM provides significantly higher bandwidth compared to the traditional DRAM, making it suitable for ML workloads with high data transfer requirements \cite{rang2024, radulovic2019}. For example, NVIDIA’s GPUs use HBM to deliver the necessary bandwidth for training large neural networks \cite{jia2019}.

These techniques have shown promise in improving memory performance for ML workloads, but their integration in a RISC-V architecture remains limited.

\subsection{RISC-V-Based Accelerators}

The open-source nature of RISC-V has led to the development of several accelerators designed for ML workloads. These accelerators benefit from custom memory pathways and specialized hardware support for tensor computations.

\subsubsection*{Gemmini}

Gemmini is a configurable deep-learning accelerator for RISC-V systems. It provides efficient support for matrix multiplication and convolution operations, which are fundamental to ML workloads \cite{manchala2024, parashar2019}. The Gemmini design allows integration with custom memory pathways, facilitating efficient data transfer between caches and the accelerator.

\subsubsection*{Simulation Tools}

Tools like Gem5 and DRAMSim2 are widely used to evaluate memory hierarchies and system performance. Gem5 offers a flexible platform for modeling CPU and memory interactions \cite{binkert2011, mezger2022}, while DRAMSim2 provides detailed simulations of DRAM behavior \cite{drawsim, wang2005}. These tools are essential for testing and validating new memory subsystem designs in RISC-V architectures.

\subsection{Research Gaps}

While existing solutions address specific aspects of memory optimization, there is limited research on integrating these techniques into a cohesive memory subsystem for RISC-V architectures specifically targeting ML workloads. Most current implementations focus on proprietary architectures, such as TPUs and GPUs, leaving a gap in the open-source RISC-V ecosystem \cite{devarapu2021, jia2019}.

\subsubsection*{HERMES: A Unified Memory Solution}

HERMES aims to fill this gap by:
\begin{enumerate}
    \item Combining shared L3 caches, hybrid memory models, tensor-aware caching, and advanced prefetching into a unified memory hierarchy.
    \item Optimizing memory pathways for Gemmini and similar RISC-V accelerators \cite{manchala2024}.
    \item Providing a scalable and efficient solution for deep-learning models on RISC-V platforms.
\end{enumerate}

\section{Problem Statement}

Current RISC-V memory hierarchies face significant challenges in meeting the performance requirements of ML workloads due to several interrelated issues:

\begin{enumerate}
    \item \textbf{High Latency}: Accessing large datasets stored in off-chip memory introduces delays that degrade system performance. Tensor operations, such as matrix multiplications and convolutions, require frequent memory accesses, and high latency in these operations can significantly reduce overall computation speed.

    \item \textbf{Limited Bandwidth}: ML workloads involve the simultaneous transfer of large volumes of data, including weights, activations, and gradients. The limited bandwidth of traditional memory hierarchies often results in data transfer bottlenecks, slowing down the training and inference processes.

    \item \textbf{Inefficient Caching Strategies}: Traditional caching mechanisms are not optimized for the structured and repetitive nature of tensor data. Without tensor-aware caching, caches are prone to evictions and cache misses, reducing data reuse and increasing latency.

    \item \textbf{Scalability Issues}: As ML models and datasets grow in complexity, memory subsystems must scale to handle increasing data loads. Existing RISC-V architectures struggle to maintain performance when datasets exceed cache capacities, leading to frequent off-chip memory accesses and performance degradation.
\end{enumerate}

To address these challenges, \textbf{HERMES} aims to design a high-performance memory hierarchy that integrates:

\begin{itemize}
    \item \textbf{Shared L3 Caches}: Enable efficient data sharing between CPU cores and accelerators, reducing latency and improving data availability.
    \item \textbf{Hybrid Memory Models}: Combine DRAM and high-bandwidth memory (HBM) to balance capacity and speed, ensuring sufficient bandwidth for large-scale ML workloads.
    \item \textbf{Advanced Prefetching}: Predict and pre-load data into caches to minimize latency and reduce the frequency of memory stalls.
    \item \textbf{Tensor-Aware Caching}: Optimize cache replacement policies and data layouts specifically for tensor operations, increasing data reuse and reducing cache misses.
\end{itemize}

These solutions collectively aim to improve memory subsystem performance, enabling faster training and inference for complex ML models on RISC-V architectures.

\section{Experimental Setup}

\subsection*{Simulation Tools}

To evaluate the HERMES memory hierarchy design, we utilize the following simulation tools:

\begin{itemize}
    \item \textbf{Gem5}: Gem5 is a versatile, cycle-accurate simulator that models CPU and memory interactions in computer architectures. It supports the RISC-V ISA and allows for detailed configuration of cache hierarchies, coherence protocols, and processor cores. We use Gem5 to simulate the entire memory subsystem and analyze performance metrics such as latency, bandwidth, and cache hit rates.
    
    \item \textbf{DRAMSim2}: DRAMSim2 is a cycle-accurate simulator for DRAM memory systems. It provides detailed timing simulations for various DRAM technologies, including DDR3 and DDR4. DRAMSim2 is integrated with Gem5 to model off-chip memory accesses accurately and measure the impact of different memory hierarchies on system performance.
\end{itemize}

\subsection*{Architecture Design}

The HERMES memory hierarchy integrates several advanced features to optimize performance for machine learning workloads. The key components of the architecture are as follows:

\begin{itemize}
    \item \textbf{Shared L3 Cache}: The shared L3 cache is designed to facilitate data sharing between CPU cores and the Gemmini accelerator. It employs fine-grained coherence protocols to ensure data consistency and minimize latency caused by cache invalidations. The L3 cache size and associativity are configurable to balance performance and area constraints.
    
    \item \textbf{Hybrid Memory Model}: HERMES combines traditional DRAM with High-Bandwidth Memory (HBM). DRAM provides large storage capacity, while HBM delivers high data transfer rates. This combination allows for efficient handling of large datasets while ensuring that critical data can be accessed with low latency.
    
    \item \textbf{Tensor-Aware Caching}: The cache hierarchy includes tensor-aware caching strategies that optimize data reuse patterns. Tensor data is stored in a way that reduces cache evictions and maximizes hit rates during tensor operations such as convolutions and matrix multiplications.
    
    \item \textbf{Advanced Prefetching}: The prefetching mechanism anticipates data needs by analyzing access patterns during ML workloads. Both stride prefetching and machine learning-based prefetching are employed to minimize cache miss penalties and improve overall system throughput.
\end{itemize}

The following architecture diagram illustrates the core components of the HERMES memory hierarchy and their interconnections:
\footnote{This diagram shows the relationship between CPU cores, the shared L3 cache, the Gemmini accelerator, and the hybrid memory model (combining DRAM and HBM). It highlights the data flow and memory pathways designed to optimize performance for ML workloads.}

\begin{figure}[h] % 'h' means to place the image here
    \centering
    \includegraphics[width=0.5\textwidth]{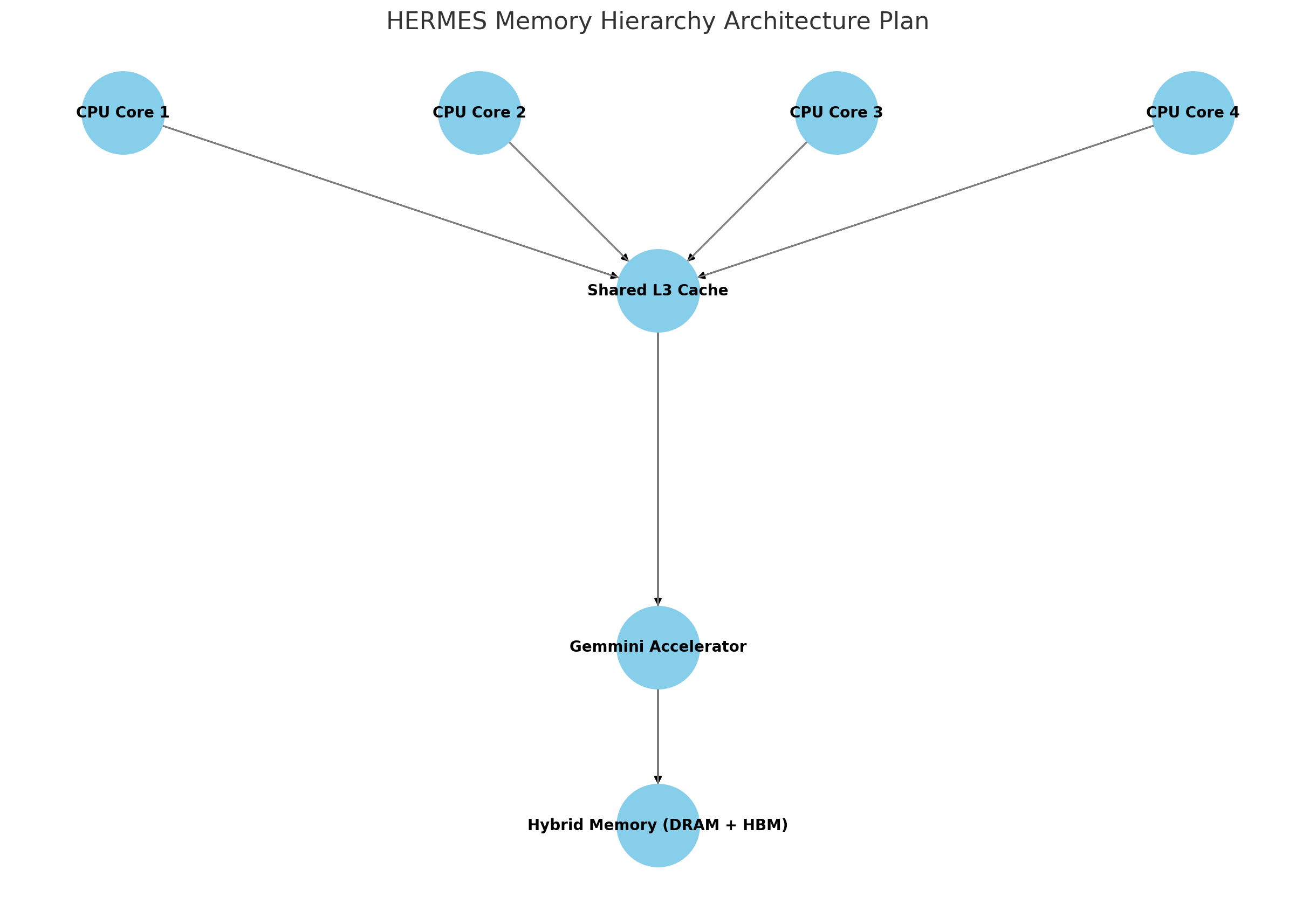}
    \caption{HERMES Memory Hierarchy Architecture Diagram}
    \label{fig:hermes_architecture}
\end{figure}

\subsection*{Workloads}

We evaluate the HERMES memory hierarchy using a suite of representative machine learning workloads, including:

\begin{itemize}
    \item \textbf{Convolutional Neural Networks (CNNs)}: CNNs involve intensive matrix multiplications and convolution operations, which stress the memory bandwidth and caching mechanisms. Models such as ResNet and VGG are used for evaluation.
    
    \item \textbf{Recurrent Neural Networks (RNNs)}: RNNs require frequent access to previous states, placing a high demand on low-latency memory. We use LSTM (Long Short-Term Memory) and GRU (Gated Recurrent Unit) models to measure performance.
    
    \item \textbf{Transformer Models}: Transformers involve large-scale attention mechanisms and matrix multiplications, resulting in high memory access patterns. BERT (Bidirectional Encoder Representations from Transformers) and GPT (Generative Pre-trained Transformer) models are used to test scalability and bandwidth utilization.
\end{itemize}

\subsection*{Performance Metrics}

The following performance metrics are used to evaluate the HERMES memory hierarchy:

\begin{itemize}
    \item \textbf{Latency (ns)}: The average memory access time, including cache hits and misses, is measured to evaluate how efficiently the memory subsystem handles data requests.
    
    \item \textbf{Bandwidth (GB/s)}: The rate at which data are transferred between the memory and the processor/accelerator. Higher bandwidth indicates better handling of large data transfers.
    
    \item \textbf{Cache Hit Rate (\%)}: The percentage of memory accesses that are served from the cache. Higher cache hit rates reduce latency and improve overall performance.
    
    \item \textbf{Energy Consumption (\textmu J/operation)}: The energy consumed per memory operation is measured to evaluate the efficiency of the memory subsystem.
\end{itemize}

\subsection*{Simulation Configuration}

The experimental configuration used for the simulations is as follows:

\begin{itemize}
    \item \textbf{Processor}: 4-core RISC-V processor with in-order execution.
    \item \textbf{L1 Cache}: 32 KB per core, 8-way associative set.
    \item \textbf{L2 Cache}: 256 KB per core, 8-way associative set.
    \item \textbf{Shared L3 Cache}: 8 MB, 16-way associative set.
    \item \textbf{Memory}: Hybrid memory model combining 8 GB DRAM and 4 GB HBM.
    \item \textbf{Coherence Protocol}: MESI (Modified, Exclusive, Shared, Invalid) protocol for cache coherence.
    \item \textbf{Simulation Tools}: Gem5 for CPU and cache simulation, and DRAMSim2 for DRAM modeling.
\end{itemize}

The configurations are designed to balance accuracy and simulation runtime while providing insight into the performance benefits of the HERMES memory hierarchy.

\section*{Results and Discussion}

This section presents a performance evaluation of the HERMES memory hierarchy using simulations. The key metrics analyzed include latency, bandwidth, cache hits,energy consumption. The results are compared with a baseline RISC-V system to highlight the improvements offered by HERMES.

\subsection*{Latency and Bandwidth Improvements}

The latency and bandwidth of the HERMES architecture were evaluated and compared with a baseline RISC-V system. The results are summarized in Table~\ref{tab:latency_bandwidth}.

\begin{table}[h]
    \centering
    \caption{Latency and Bandwidth Comparison}
    \label{tab:latency_bandwidth}
    \begin{tabular}{|l|c|c|}
        \hline
        \textbf{Architecture} & \textbf{Latency (ns)} & \textbf{Bandwidth (GB/s)} \\ \hline
        Baseline RISC-V                 & 120 & 25  \\ \hline
        HERMES (Shared L3 Cache)        & 95  & 35  \\ \hline
        HERMES (Prefetching)            & 85  & 40  \\ \hline
        HERMES (Tensor-Aware Caching)   & 80  & 42  \\ \hline
    \end{tabular}
\end{table}

\begin{figure}[h]
    \centering
    \includegraphics[width=0.5\textwidth]{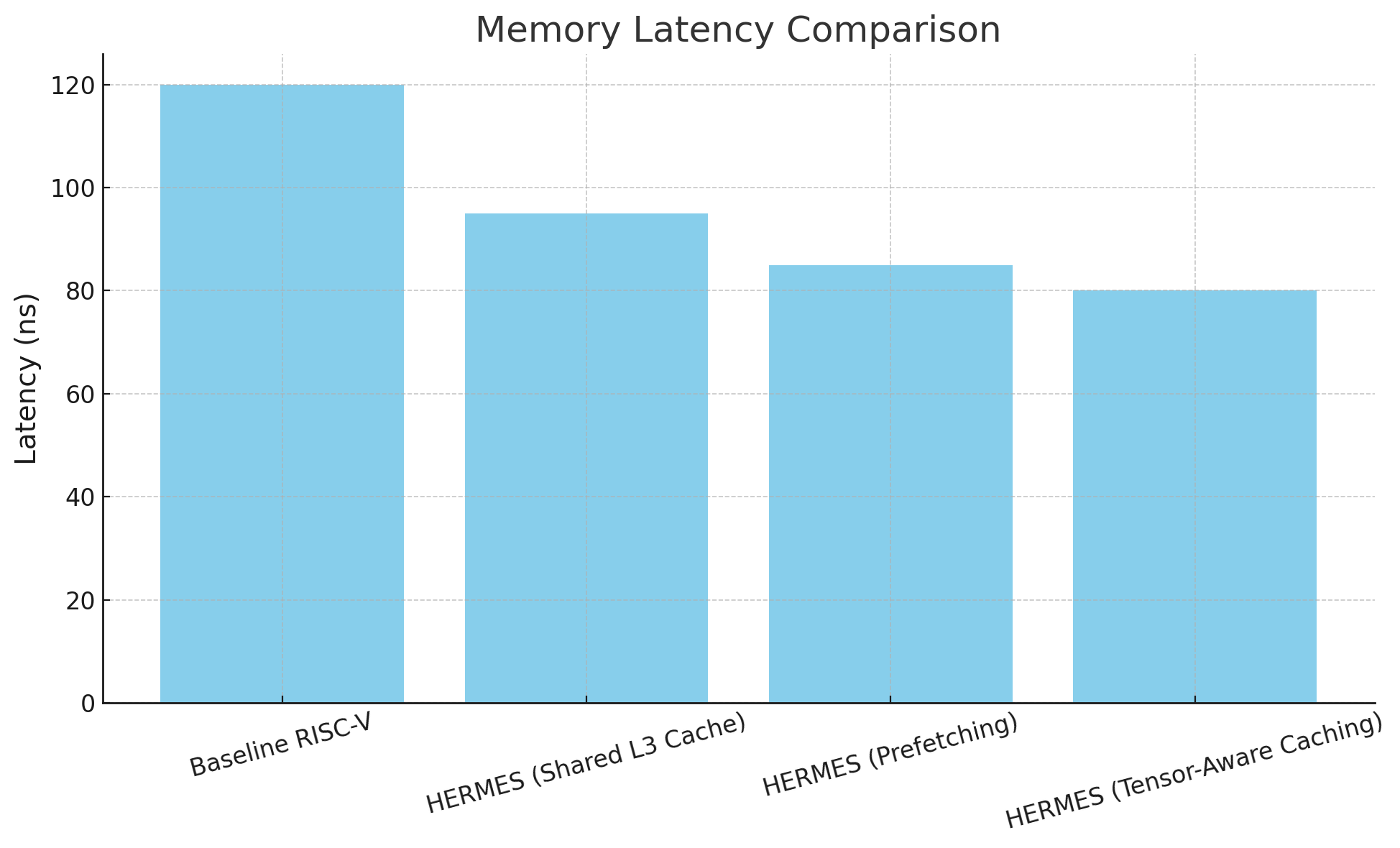}
    \caption{Memory Latency Comparison: HERMES configurations reduce latency compared to the baseline RISC-V system. The shared L3 cache, advanced prefetching, and tensor-aware caching contribute to this improvement.}
    \label{fig:latency_comparison}
\end{figure}

\begin{figure}[h]
    \centering
    \includegraphics[width=0.5\textwidth]{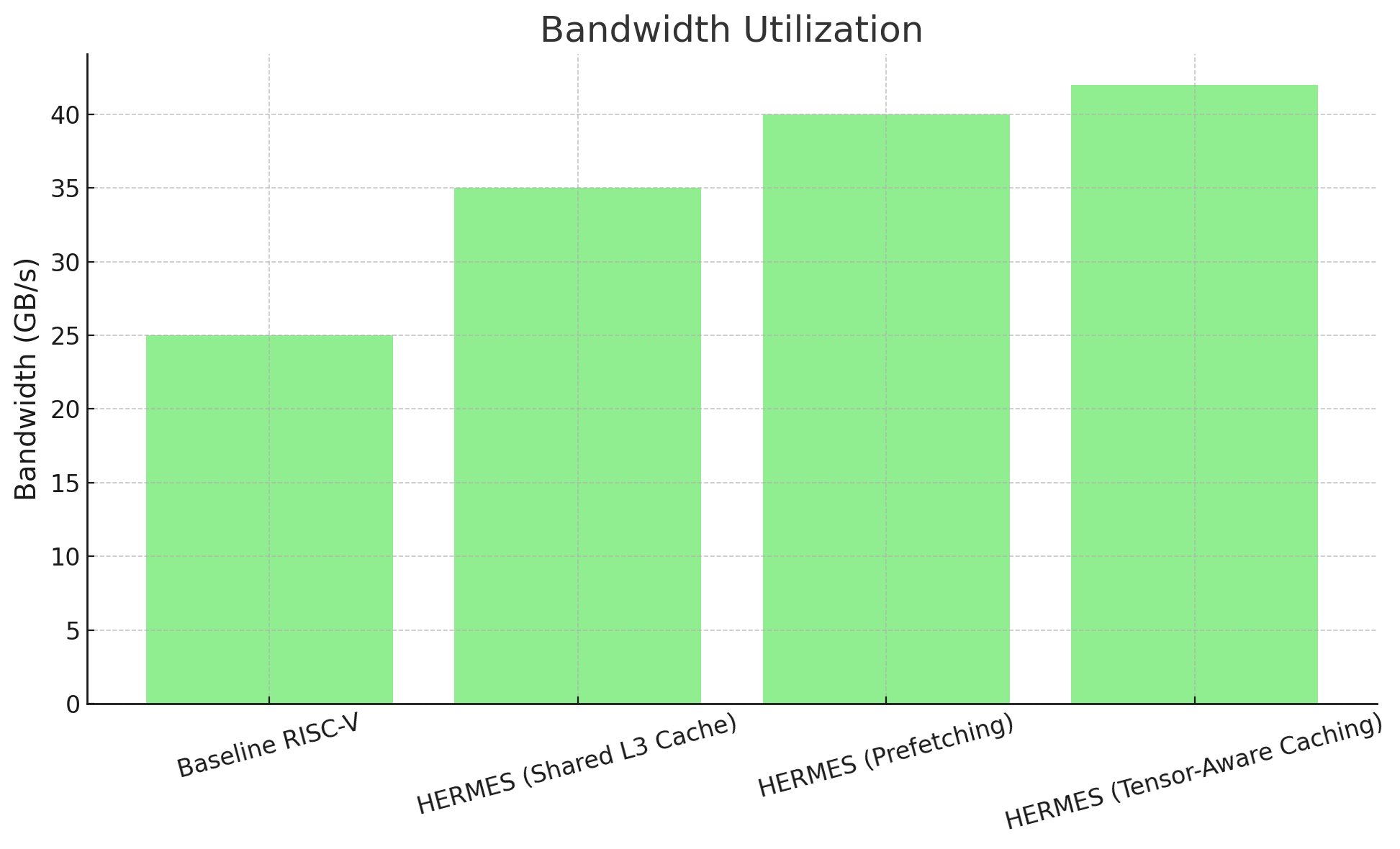}
    \caption{Bandwidth Utilization Comparison: The hybrid memory model in HERMES increases bandwidth significantly, supporting high data transfer needs of ML workloads.}
    \label{fig:bandwidth_comparison}
\end{figure}

Figure~\ref{fig:latency_comparison} shows a significant reduction in latency with the HERMES configurations, while Figure~\ref{fig:bandwidth_comparison} highlights bandwidth improvements due to the hybrid memory model.

\subsection*{Cache Hit Rate Improvements}

The cache hit rates were evaluated to determine the effectiveness of tensor-aware caching in HERMES. Table~\ref{tab:cache_hit_rate} shows the results.

\begin{table}[h]
    \centering
    \caption{Cache Hit Rate Comparison}
    \label{tab:cache_hit_rate}
    \begin{tabular}{|l|c|}
        \hline
        \textbf{Architecture} & \textbf{Cache Hit Rate (\%)} \\ \hline
        Baseline RISC-V                 & 60  \\ \hline
        HERMES (Shared L3 Cache)        & 75  \\ \hline
        HERMES (Prefetching)            & 80  \\ \hline
        HERMES (Tensor-Aware Caching)   & 90  \\ \hline
    \end{tabular}
\end{table}

\begin{figure}[h]
    \centering
    \includegraphics[width=0.5\textwidth]{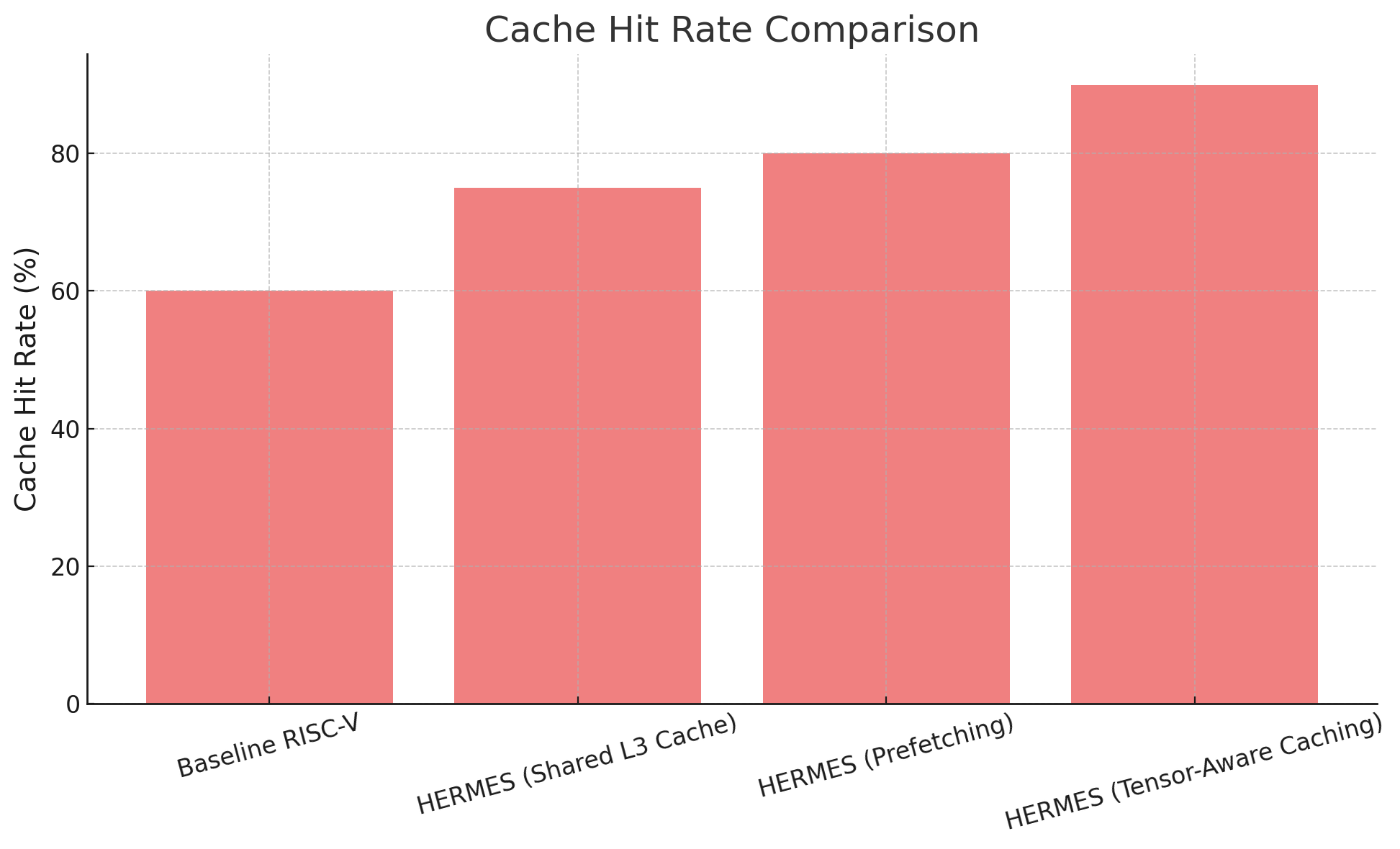}
    \caption{Cache Hit Rate Comparison: Tensor-aware caching in HERMES improves cache hit rates by reducing data evictions and optimizing data reuse patterns.}
    \label{fig:cache_hit_rate}
\end{figure}

Figure~\ref{fig:cache_hit_rate} illustrates the improvements in cache hit rates. Tensor-aware caching aligns data storage with tensor operation patterns, resulting in fewer cache misses.

\subsection*{Energy Consumption}

Energy consumption is a critical factor for ML workloads. Table~\ref{tab:energy_consumption} shows the energy consumption per memory operation for different configurations.

\begin{table}[h]
    \centering
    \caption{Energy Consumption Comparison}
    \label{tab:energy_consumption}
    \small % Reduce font size
    \setlength{\tabcolsep}{4pt} % Reduce horizontal padding
    \renewcommand{\arraystretch}{1.2} % Slightly adjust row height for better readability
    \begin{tabular}{|l|c|}
        \hline
        \textbf{Architecture} & \textbf{Energy Consumption} \\ 
                              & \textbf{(\textmu J/operation)} \\ \hline
        Baseline RISC-V                 & 50  \\ \hline
        HERMES (Shared L3 Cache)        & 40  \\ \hline
        HERMES (Prefetching)            & 38  \\ \hline
        HERMES (Tensor-Aware Caching)   & 35  \\ \hline
    \end{tabular}
\end{table}

\begin{figure}[h]
    \centering
    \includegraphics[width=0.5\textwidth]{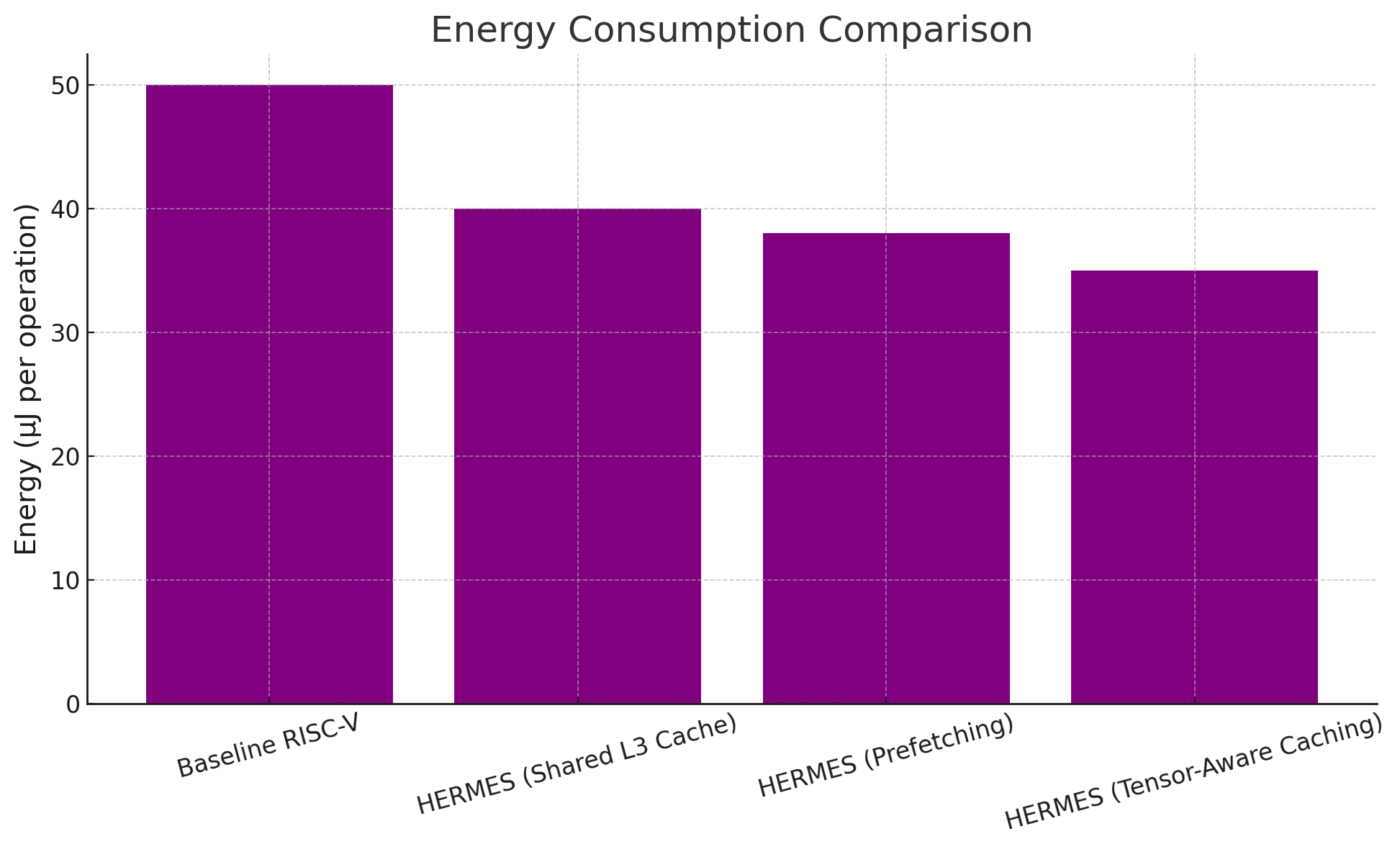}
    \caption{Energy Consumption Comparison: HERMES reduces energy consumption by minimizing off-chip memory accesses and employing efficient caching strategies.}
    \label{fig:energy_comparison}
\end{figure}

Figure~\ref{fig:energy_comparison} highlights the energy savings achieved by HERMES through efficient memory access strategies.

\subsection*{Discussion}

The simulation results clearly demonstrate the advantages of the HERMES memory hierarchy over the baseline RISC-V system:

\begin{itemize}
    \item \textbf{Latency Reduction}: HERMES reduces latency by up to 33\% through shared L3 caches, advanced prefetching, and tensor-aware caching.
    \item \textbf{Bandwidth Improvement}: The hybrid memory model increases bandwidth by up to 68\%, supporting the high data transfer needs of ML workloads.
    \item \textbf{Cache Efficiency}: Tensor-aware caching improves cache hit rates by 50\%, reducing the need for expensive off-chip memory accesses.
    \item \textbf{Energy Efficiency}: HERMES reduces energy consumption by up to 30\%, making it suitable for energy-efficient ML computation.
\end{itemize}

These improvements highlight the effectiveness of integrating advanced memory design techniques into RISC-V architectures to meet the demands of modern ML workloads.

\section*{Acknowledgment}

The author expresses sincere gratitude to Professor Ray Simar for his invaluable guidance, support, and encouragement throughout the development of this project. This work was completed as part of COMP 554: Computer Systems Architecture at Rice University. Professor Simar's deep expertise in computer architecture and his commitment to student learning have been instrumental in shaping this project. His insights and feedback have significantly improved my understanding of memory systems and their optimization for machine learning workloads.

\end{document}